\documentclass[twocolumn,prl,showpacs,preprintnumbers,amsmath,amssymb,superscriptaddress]{revtex4}
\usepackage{graphicx,color}
\usepackage{amssymb}
\usepackage{amsmath}
\usepackage{latexsym}
\usepackage{amsxtra}
\usepackage{amscd}
\usepackage{dcolumn}
\usepackage{bm}
\usepackage{hyperref}
\graphicspath{}
\let\oldsqrt\sqrt 
\def\sqrt{\mathpalette\DHLhksqrt}
\def\DHLhksqrt#1#2{\setbox0=\hbox{$#1\oldsqrt{#2\,}$}\dimen0=\ht0
\advance\dimen0-0.2\ht0
\setbox2=\hbox{\vrule height\ht0 depth -\dimen0}%
{\box0\lower0.4pt\box2}}
\begin{document}

\title{Correlations in intermediate-energy two-proton removal reactions}

\author{K.~Wimmer}\altaffiliation[Present address ]{Central Michigan University, Mt. Pleasant, Michigan 48859, USA}
\author{D.~Bazin}
\affiliation{National Superconducting Cyclotron Laboratory, Michigan State University, East Lansing, Michigan 48824, USA}

\author{A.~Gade}
\affiliation{National Superconducting Cyclotron Laboratory, Michigan State University, East Lansing, Michigan 48824, USA}
\affiliation{Department of Physics and Astronomy, Michigan State University, East Lansing, Michigan 48824, USA}

\author{J.~A.~Tostevin}
\affiliation{National Superconducting Cyclotron Laboratory, Michigan State University, East Lansing, Michigan 48824, USA}
\affiliation{Department of Physics, University of Surrey, Guildford, Surrey GU2 7XH, United Kingdom}

\author{T.~Baugher}
\affiliation{National Superconducting Cyclotron Laboratory, Michigan State University, East Lansing, Michigan 48824, USA}
\affiliation{Department of Physics and Astronomy, Michigan State University, East Lansing, Michigan 48824, USA}

\author{Z.~Chajecki}
\affiliation{National Superconducting Cyclotron Laboratory, Michigan State University, East Lansing, Michigan 48824, USA}

\author{D.~Coupland}
\affiliation{National Superconducting Cyclotron Laboratory, Michigan State University, East Lansing, Michigan 48824, USA}
\affiliation{Department of Physics and Astronomy, Michigan State University, East Lansing, Michigan 48824, USA}

\author{M.~A.~Famiano}
\affiliation{Department of Physics, Western Michigan University, Kalamazoo, Michigan 49008, USA}

\author{T.~K.~Ghosh}
\affiliation{Variable Energy Cyclotron Centre, 1/AF Bidhannagar, Kolkata 700064, India}

\author{G.~F.~Grinyer}\altaffiliation[Present address ]{GANIL, CEA/DSM-CNRS/IN2P3, Bvd Henri Becquerel, 14076 Caen, France}
\affiliation{National Superconducting Cyclotron Laboratory, Michigan State University, East Lansing, Michigan 48824, USA}

\author{R.~Hodges}
\affiliation{National Superconducting Cyclotron Laboratory, Michigan State University, East Lansing, Michigan 48824, USA}
\affiliation{Department of Physics and Astronomy, Michigan State University, East Lansing, Michigan 48824, USA}

\author{M.~E.~Howard}
\affiliation{Department of Physics and Astronomy, Rutgers University, New Brunswick, New Jersey 08903, USA}

\author{M.~Kilburn}
\author{W.~G.~Lynch}
\affiliation{National Superconducting Cyclotron Laboratory, Michigan State University, East Lansing, Michigan 48824, USA}
\affiliation{Department of Physics and Astronomy, Michigan State University, East Lansing, Michigan 48824, USA}

\author{B.~Manning}
\affiliation{Department of Physics and Astronomy, Rutgers University, New Brunswick, New Jersey 08903, USA}

\author{K.~Meierbachtol}
\affiliation{National Superconducting Cyclotron Laboratory, Michigan State University, East Lansing, Michigan 48824, USA}
\affiliation{Department of Chemistry, Michigan State University, East Lansing, Michigan 48824, USA}

\author{P.~Quarterman}
\author{A.~Ratkiewicz}
\author{A.~Sanetullaev}
\affiliation{National Superconducting Cyclotron Laboratory, Michigan State University, East Lansing, Michigan 48824, USA}
\affiliation{Department of Physics and Astronomy, Michigan State University, East Lansing, Michigan 48824, USA}

\author{E.~C.~ Simpson}
\affiliation{Department of Physics, University of Surrey, Guildford, Surrey GU2 7XH, United Kingdom}

\author{S.~R.~Stroberg}
\affiliation{National Superconducting Cyclotron Laboratory, Michigan State University, East Lansing, Michigan 48824, USA}
\affiliation{Department of Physics and Astronomy, Michigan State University, East Lansing, Michigan 48824, USA}

\author{M.~B.~Tsang}
\affiliation{National Superconducting Cyclotron Laboratory, Michigan State University, East Lansing, Michigan 48824, USA}

\author{D.~Weisshaar}
\affiliation{National Superconducting Cyclotron Laboratory, Michigan State University, East Lansing, Michigan 48824, USA}

\author{J.~Winkelbauer}
\affiliation{National Superconducting Cyclotron Laboratory, Michigan State University, East Lansing, Michigan 48824, USA}
\affiliation{Department of Physics and Astronomy, Michigan State University, East Lansing, Michigan 48824, USA}

\author{R.~Winkler}
\affiliation{National Superconducting Cyclotron Laboratory, Michigan State University, East Lansing, Michigan 48824, USA}

\author{M.~Youngs}
\affiliation{National Superconducting Cyclotron Laboratory, Michigan State University, East Lansing, Michigan 48824, USA}
\affiliation{Department of Physics and Astronomy, Michigan State University, East Lansing, Michigan 48824, USA}

\begin{abstract}
We report final-state-exclusive measurements of the light charged fragments in
coincidence with $^{26}$Ne residual nuclei following the direct two-proton removal
from a neutron-rich $^{28}$Mg secondary beam. A Dalitz-plot analysis and comparisons
with simulations show that a majority of the triple-coincidence events with two
protons display phase-space correlations consistent with the (two-body) kinematics
of a spatially-correlated pair-removal mechanism. The fraction of such correlated
events, 56(12)~\%, is consistent with the fraction of the calculated cross section,
64~\%, arising from spin $S=0$ two-proton configurations in the entrance-channel
(shell-model) $^{28}$Mg ground state wave function. This result promises access
to an additional and more specific probe of the spin and spatial correlations of valence
nucleon pairs in exotic nuclei produced as fast secondary beams.
\end{abstract}

\date{\today}

\maketitle

Access to nuclear reactions that can probe the states of pairs of nucleons
in an atomic nucleus is a long-standing ambition. Specifically, an ability
to probe the spin-structure of nucleon pairs and, for example, to identify
and quantify two-nucleon correlations in the $[S,T]$=$[0,1]$ spin-isospin
channel, is required. Intermediate-energy reactions that remove two nucleons
(2N) suddenly from a fast projectile, in collisions with a light target
nucleus, have been shown~\cite{tostevin06,simpson10,simpson09,simpson09b}
to provide a sensitivity to the 2N configurations near the projectile
surface. Experimentally, the importance of nucleon removal reactions derives
from their high detection efficiency (forward focusing of the reaction
residues) and use of thick reaction targets, increasing the effective
luminosity of the relatively low intensity exotic beams. Prior to the
present work these 2N removal cross section measurements have been
inclusive with respect to the final states of both the target and the
removed nucleons but often exclusive with respect to the bound final-states
of the forward-traveling projectile-like residues, allowing spectroscopic
studies based on the different residue final-state yields and the shapes
of their momentum distributions \cite{Yoneda,FSU1,Bastin,LBL1,FSU3,bazin03}.

The present work sacrifices, temporarily, this better-understood residue
final-state sensitivity to investigate new information that might be forthcoming
from more exclusive measurements of the final states of the removed nucleons.
The $^{9}$Be($^{28}$Mg,$^{26}$Ne) reaction was used at an intermediate energy
of 93 MeV/u. The chosen reaction was studied previously in a $^{26}$Ne-$\gamma$
coincidence measurement~\cite{bazin03} and was used to confirm the predicted
relative populations of the four bound $^{26}$Ne final states~\cite{tostevin06}.
The present data set has also been used~\cite{wimmer12} to confirm that the
measured contributions to the inclusive 2N-removal cross section from each
of the possible elastic and inelastic removal mechanisms~\cite{tostevin06}
were consistent with calculations that use eikonal reaction dynamics and
$sd$-shell-model structure inputs for the $^{28}$Mg to $^{26}$Ne$(J^{\pi})$
2p overlap functions. There it was shown that only $8(2)$~\% of the inclusive
$^{9}$Be($^{28}$Mg,$^{26}$Ne) reaction cross section results from elastic
breakup, the remainder being associated with reactions in which at least
one of the removed protons interacts inelastically with the $^{9}$Be target
nucleus~\cite{wimmer12}.

The experiment was carried out at the Coupled Cyclotron Facility at the
NSCL(MSU). The $^{28}$Mg secondary beam was produced by projectile
fragmentation of a 140 MeV/u $^{40}$Ar primary beam and selected using
the A1900 fragment separator~\cite{morrissey03}. The $^{9}$Be reaction
target had a thickness of 100~mg/cm$^2$ and was placed at the target
position of the high-resolution S800 magnetic spectrograph~\cite{bazins80003}.
The beam energy at mid-target position was 93~MeV/u. The $^{26}$Ne
reaction residues were identified by measuring the energy loss in the
ionization chamber in the S800 focal plane and the time of flight measured
between scintillators before and after the reaction target. The $^{26}$Ne
energy and momenta were reconstructed from the measured positions and
angles in the focal plane using the position-sensitive cathode readout
drift chambers (CRDC) and ion optical ray-tracing. Protons and other
light charged particles were detected and identified in the high-resolution
array HiRA~\cite{wallace07}. Other experimental details and the angular
coverage of HiRA were discussed in detail in Ref.~\cite{wimmer12}. In the
present analysis we considered those triple-coincidence events where both
of the light particles were identified as protons. These 4810 events
represent 28(3)~\% of the inclusive two-proton removal cross
section~\cite{wimmer12}.

\begin{figure}[h]
\centering
\includegraphics[width=\columnwidth]{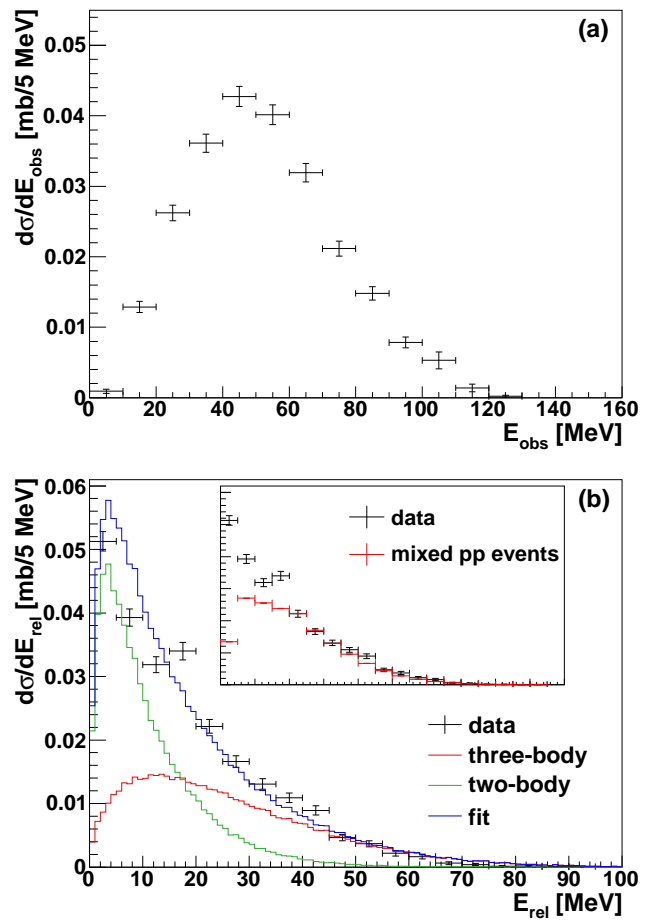}
\caption{(color online) (a) Energy of the dissociated $^{26}$Ne+p+p fragments,
in their $P_{c12}=0$ frame. The cross section has been corrected for the
geometric efficiency of the HiRA array, as described in~\cite{wimmer12}.
(b) Relative energy distribution of the two detected protons. The fractions
of simulated three-body (red) and two-body (green) kinematics events have
been fitted to the data set (see text for details). The inset compares the
measured relative energy distribution with that derived from mixing protons from two independent events (red). The distribution obtained from 
mixed events has been scaled to the experimental two-proton relative energy 
distribution above 40~MeV} \label{fig:decrel}
\end{figure}

The continuum energy of these $^{26}$Ne+p+p triple-coincidence events
in their total momentum $P_{c12}=0$ frame can be reconstructed from the
three measured laboratory frame four-momenta $P_\text{i}$. Thus,
\begin{equation}
E_\text{obs} = \sqrt{ (\sum P_\text{i})^2 } - \sum m_\text{i}~
\end{equation}
is the energy (above threshold) of the dissociated $^{28}$Mg fragments,
whose distribution is shown in Fig.~\ref{fig:decrel}(a). This distribution
served as an input for Monte-Carlo kinematics simulations, used to
populate the $^{26}$Ne+p+p phase space and explore correlations of
the detected protons.

In the simulations the three particles were assumed to be structure-less.
There was no consideration of spin degrees of freedom of the two protons.
Two kinematics scenarios were used, (a) three-body and (b) two-body, to
generate the four-momenta of the three particles in the $P_{c12}=0$ frame. The summed
energies of the three particles were sampled from the experimental
$d\sigma/dE_\text{obs}$ distribution shown in Fig. \ref{fig:decrel}(a).
In the three-body case the available energy $E_\text{obs}$ was distributed
among the three particles (democratic breakup). The residue and protons
are then correlated only by energy and momentum conservation leading to
the broad two-proton relative energy distribution shown in Fig.
\ref{fig:decrel}(b). The two-body case simulated events where the target
was assumed to have encountered spatially-localized proton pairs (carrying
energy $\varepsilon^*$ in their $P_{12}=0$ frame) and delivered an impulse
to these pairs; as would be expected for some fraction of events in the
2N-removal reaction mechanism that favors the surface localization and
spatial proximity of pairs of nucleons (see e.g.~\cite{simpson09b}). The
energy $E_\text{obs}$ was now shared between an assumed $\varepsilon^*$
and the motion of the center-of-mass of the protons, producing $^{26}$Ne+(2p)*
configurations that decay to $^{26}$Ne+p+p.

To study the two-proton correlations we first reconstruct the measured
two-proton relative energy, $E_\text{rel}$, distribution shown in Fig.
\ref{fig:decrel}(b). These data points were fitted to extract the relative
numbers of three-body (red) and two-body (green) kinematics events, the events 
for $E_\text{rel}>40$~MeV being assumed to result from the three-body mechanism 
only. The fitted two-body events then determine the $\varepsilon^*$ spectrum 
that is sampled to generate the two-body phase-space.
We attribute the fraction 0.56(12) to these latter, correlated 2p-removal events.
The two kinematics scenarios also show a different signature with respect to
the angle between the protons. However, due to both the coarse and somewhat
limited azimuthal angles coverage in the experiment, no additional useful
evidence could be extracted from the measured angular distributions. We
note that an $E_\text{rel}$ distribution generated from mixed events --
in which the two protons were
completely uncorrelated -- did not show the increase in the cross section
for small $E_\text{rel}$, as is shown in the inset to Fig.~\ref{fig:decrel}(b).

\begin{figure}[h]
\centering
\includegraphics[width=\columnwidth]{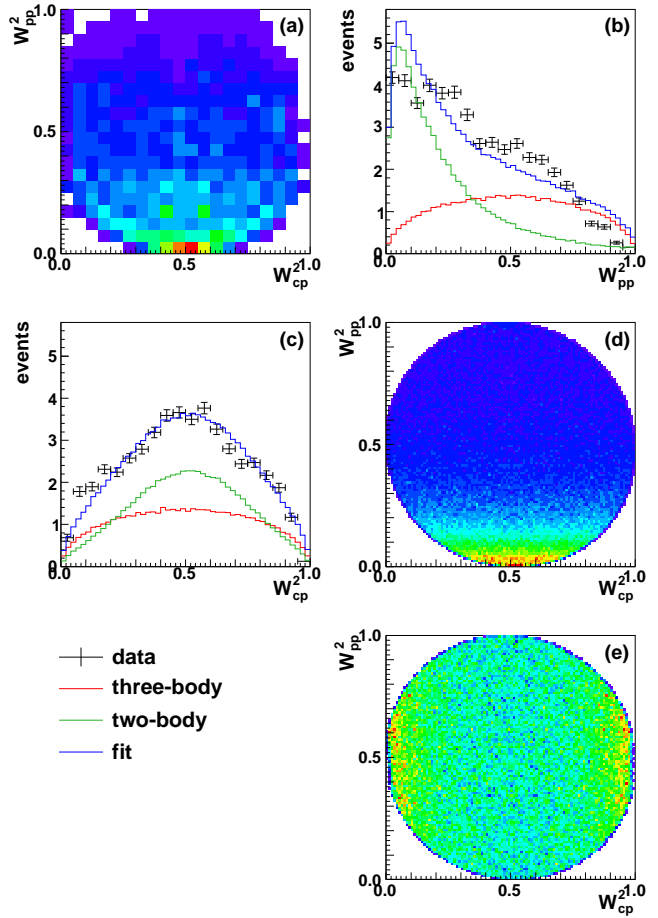}
\caption{(color online) Dalitz plots of core ($^{26}$Ne residue)-proton
$W_\text{cp}$ and proton-proton $W_\text{pp}$ invariant masses. Panel (a)
shows the experimental data with projections on $W^2_\text{pp}$ (b) and
$W^2_\text{cp}$ (c). The results of the simulation using a two-body fraction 
of 0.56 are shown in panel (d). Panel (e) shows the corresponding
Dalitz plot for simulated sequential two-proton removal events, via a
single-proton removal to a proton-unbound intermediate-state in $^{27}$Na.}
\label{fig:dalitz}
\end{figure}

Sequential two-proton removal events, via a single-proton removal to
proton-unbound intermediate-states in $^{27}$Na, are not considered to
contribute. Such indirect paths would involve the proton decay of
intermediate $^{27}$Na excited states with energies in excess of 13.3
MeV, whereas the neutron threshold in $^{27}$Na lies below 7 MeV in
excitation energy~\cite{bazin03}. This expectation is confirmed
experimentally by the analysis of the Dalitz plot in Fig.~\ref{fig:dalitz},
as will now be discussed.

In high-energy physics, three-particle final-states are commonly studied by
the analysis of Dalitz plots~\cite{dalitz54}. This shows the correlation of
the squares of the invariant masses $M^2_{ij} = (P_i+P_j)^2$. In the absence
of sources of two-particle correlations, the observed distributions will be
uniform within the boundaries defined by energy and momentum conservation.
Since the quantity on the Dalitz plot is proportional to the square of the
matrix element for the system, intermediate states as well as symmetries
generate a non-uniform distribution. In the present case, the boundaries
will be different for each value of the continuum energy, $E_\text{obs}$,
so we have adopted the approach of Ref.~\cite{marques01} and used the
normalized invariant masses, $W_{ij}$, that range from 0 to 1, i.e.
\begin{equation}
W^2_{ij} = \frac{M^2_{ij}-(m_i+m_j)^2}{(E_\text{obs}+m_i+m_j)^2-(m_i+m_j)^2}~.
\end{equation}

The Dalitz plot of the current $^{26}$Ne+p+p event data is shown in Fig.
\ref{fig:dalitz}(a).
The experimental results are compared to the three-body and two-body
2p-removal kinematics simulations using a two-body fraction of 0.56 in panel (d). The three-body
case results in a flat distribution while the two-body kinematics
simulation produces a non-uniform filling of the phase space -- with an
increased intensity at lower proton-proton invariant masses $W_\text{pp}$.

If present, sequential two-proton removal via an excited intermediate state
in $^{27}$Na would lead to vertical bands with constant $W^2_\text{cp}$,
as is shown in panel (e). This was not observed in the experimental data,
providing strong support for our assumption that such indirect paths do
not contribute. This provides a direct experimental confirmation of earlier
arguments, based on energetics, the cross section, and the inclusive
momentum distribution, for the dominance of the direct removal of the
two strongly-bound valence particles~\cite{bazin03,tostevin06}.

Also shown, in panels (b) and (c) of Fig.~\ref{fig:dalitz}, are the data
projections onto the $W^2_\text{pp}$ and $W^2_\text{cp}$ axes. These
data projections are also reasonably well described by the combination of the two breakup
modes determined from the fit to the relative energy spectrum, Fig.~\ref{fig:decrel}(b).
We note that the effect of the $^{9}$Be
reaction target enters the simulations only very indirectly, through
the measured projectile excitation energy distribution, $d\sigma/
dE_\text{obs}$ of Fig.~\ref{fig:decrel}(a). Thus, the comparisons
of the data with the two simulated kinematics scenarios does not
provide information on the final states of the target, to which our
results remain inclusive.

It is instructive to compare the deduced correlated 2p-fraction
to the calculated two-nucleon spin contributions to the reaction cross
sections. The partial cross sections to each $^{26}\text{Ne}(J^\pi)$
final state receive incoherent (additive) contributions from the total
orbital $L$ and spin $S$ angular momentum components in the $\langle
^{26}\text{Ne}(J^\pi)|^{28}\text{Mg} \rangle$ two-proton overlaps
\cite{simpson10}. Table~\ref{table1} shows the $S$=0 fractions (\%)
of both the overlaps and the removal cross sections, $\sigma_{S=0}$,
when using the $sd$-shell-model two-nucleon amplitudes tabulated
in Ref.~\cite{tostevin04}. These $[S,T]$=$[0,1]$ wave function
components will have significant but not exclusively $^1\!s_0$
2p configurations.
\begin{table}[h]
\caption{Calculated partial and inclusive cross sections for the
$^{28}$Mg($-$2p) reaction at 93 MeV per nucleon and the percentage
contributions of $S$=0 terms to the two-proton overlaps and the
cross sections. The 2p stripping contributions are shown in Table
III of Ref. \cite{tostevin04}, there for 82.3MeV/nucleon.
\label{table1}}
\begin{ruledtabular}
\begin{tabular}{lccccc}
$J_f^\pi$&$E$&Overlap &
~~$\sigma_{th}$~~ &~~$\sigma_{S=0}$~~&~~$\sigma_{S=0}$~~\\
&(MeV)&($S=0$, \%)& (mb) & (mb) & (\%)\\
\hline
0$^+$&0.0 &86&1.190 &1.083&90\\
2$^+$&2.02&18&0.327 &0.071&22\\
4$^+$&3.50&38&1.046 &0.523&49\\
$2_2^+$&3.70&50&0.458 &0.250&54\\
\hline
Incl.  &&    &3.02  &1.93&64\\
\end{tabular}
\end{ruledtabular}
\end{table}

The relative transparency of the direct reaction mechanism to the $S$
of the two nucleons delivered by the projectile is evident from
Table~\ref{table1}. The $S=0$ fractions of the overlaps are reflected
rather directly, with only small final-state-dependent enhancements,
in their percentage contributions to the cross sections,
$\sigma_{S=0}$. Here the $S$=0 terms are seen to be responsible for
64~\% of the computed inclusive cross section, consistent with the
two-body event fraction deduced from the simulations. This suggests
that the $[S,T]$=$[0,1]$ correlations present in that part of the
$^{28}$Mg ground-state wave function sampled by the target may be
maintained in the sudden two-proton removal process and be reflected,
here via the two-body kinematics events simulation, as two-proton
correlations in the final state.

\begin{figure}[h]
\centering
\includegraphics[width=\columnwidth]{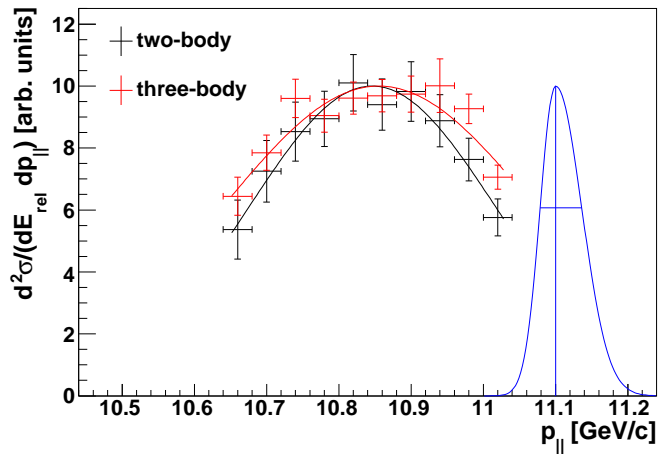}
\caption{(color online) Parallel momentum distribution of the $^{26}$Ne
residue gated on two-body (black) and three-body (red) events. The two 
components were extracted by applying a cut on $E_\text{rel}$ at the 
intersection of the two distributions in Fig.~\ref{fig:decrel}(b). 
Contributions from three-body events at low $E_\text{rel}$ were 
subtracted from the two-body distribution and vice-versa for the two-body 
events with high relative energy.
The lines are Gaussian fits. The two distributions have
been normalized to show the difference in width. The blue distribution
illustrates the width of the momentum distribution of the incoming
$^{28}$Mg beam.} \label{fig:erelcut}
\end{figure}

This being the case, an expectation is that the width of the parallel
momentum distribution of residues, gated on the two-body-like events
with smaller $E_\text{rel}$, should be narrower; this due to the size
of the $^{26}$Ne($0^+$) ground state cross section with its narrow
momentum distribution~\cite{simpson09}. However, as all $^{26}$Ne final
states contribute significantly to the 64~\% $S=0$ contribution to the
inclusive cross section, this differential width is only slightly
larger than our current experimental resolution. Fig.~\ref{fig:erelcut}
shows the parallel momentum distribution of the $^{26}$Ne residues (in
coincidence with two protons) gated on two-body and three-body events. We only consider the central part of momentum
distributions since the high and low-momentum tails were influenced by
acceptance cut-offs and proton detection thresholds.
The parallel momentum distribution gated on three-body mechanism events (red 
line in Fig.~\ref{fig:erelcut}) is slightly wider than the corresponding
distribution gated on two-body events. This differential width effect is
consistent with what is expected based on the theoretical $S=0$ and $S=1$
momentum distributions folded with the width of the momentum distribution 
of the unreacted $^{28}$Mg beam. Since the two- and three-body simulations 
describe the available data, and the extracted momentum distributions are 
consistent, within statistics, with the two-body events being closely 
allied with the $S=0$ overlap function components, we may expect that the 
less-spatially-localized $S=1$ two-proton components will track the
results of the three-body simulation. Additional data, e.g. to states with 
different $S=0,1$ fractions will be needed to confirm this expectation.

In summary, we have observed significant kinematical correlations of
two final-state protons measured in coincidence with $^{26}$Ne residues
following the $^{9}$Be($^{28}$Mg,$^{26}$Ne) two-proton removal reaction.
We attribute the deduced 56(12)~\% fraction of two-body events to the
observed cross section as due to spin $S=0$ two-proton configurations
in the $^{28}$Mg wave function in the entrance-channel. This result
suggests the potential for such measurements to provide an additional,
more specific probe of the spin correlations of valence nucleon pairs
in exotic nuclei. It also suggests the need for additional exclusive
measurements to confirm and quantify this proposed spin sensitivity.
Specifically, two-nucleon removal reactions to nuclei with only a
single ($0^+$) bound state, where $S=0$ configurations will dominate,
would be ideally suited for this kind of study. Further indirect
indications of $S=0$ driven final-state 2p correlations could come
from observations of enhanced $^3$He and $\alpha$-particle yields
produced following 1n- or 2n-pickup from the target by $S=0$ 2p
configurations in their surface-grazing removal collisions.

This work was supported by the National Science Foundation under
Grants Nos. PHY-0606007, PHY-0757678 and PHY-1102511, and the UK
Science and Technology Facilities Council (STFC) under Grant No.
ST/J000051/1.

\bibliography{correlation}

\end{document}